\documentclass[twocolumn,showpacs,amsmath,amssymb,pre]{revtex4-1}

\usepackage{graphicx}
\usepackage{color}
\usepackage{dcolumn}
\usepackage{amsmath}
\usepackage{CJK}
\usepackage{subfigure}

\usepackage{sidecap}
\usepackage{dsfont}

\usepackage{mathrsfs}
\usepackage{amsfonts}
\usepackage{indentfirst}
\usepackage{bm}

\usepackage[colorlinks,citecolor=blue]{hyperref}

\newcommand{\be}{\begin{equation}}
\newcommand{\ee}{\end{equation}}
\newcommand{\bey}{\begin{eqnarray}}
\newcommand{\eey}{\end{eqnarray}}
\newcommand{\bw}{\begin{widetext}}
\newcommand{\ew}{\end{widetext}}

\newcommand{\ra}{\rangle}
\newcommand{\la}{\langle}
\newcommand{\bq}{ {\bf q} }
\newcommand{\bp}{ {\bf p} }

\newcommand{\ov}{\overline }

\newcommand{\ba}{\begin{array}}
\newcommand{\ea}{\end{array}}
\newcommand{\bi}{\begin{itemize}}
\newcommand{\ei}{\end{itemize}}
\newcommand{\bem}{\begin{enumerate}}
\newcommand{\eem}{\end{enumerate}}

\begin{document}

\title{Signatures of excited state quantum phase transitions in quantum many body systems: Phase space analysis}

\author{Qian Wang}
\affiliation{Department of Physics, Zhejiang Normal University, Jinhua 321004, China, \\
CAMTP-Center for Applied Mathematics and Theoretical Physics, University of Maribor, 
Mladinska 3, SI-2000 Maribor, Slovenia}

\author{Francisco P\'{e}rez-Bernal}
\affiliation{Departamento de Ciencias Integradas y Centro de Estudios Avanzados en F\'{i}sica,
Matem\'{a}ticas y Computaci\'{o}n, Universidad de Huelva, Huelva 21071, Spain and Instituto Carlos
I de F\'{i}sica Te\'{o}rica y Computacional, Universidad de Granada, Granada 18071, Spain}

\begin{abstract}

Using the Husimi function, we investigate the phase space signatures of the excited state quantum phase transitions
(ESQPTs) in the Lipkin and coupled top models.
We show that the time evolution of the Husimi function exhibits distinct behaviors 
between the different phases of an ESQPT and the presence of an ESQPT is signaled by the 
particular dynamics of the Husimi function.
We also evaluate the long time averaged Husimi function and its associated marginal distributions, 
and discuss how to identify the signatures of ESQPT from their properties.
Moreover, by exploiting the second moment and Wherl entropy of the long-time averaged Husimi function, 
we estimate the critical points of ESQPTs, demonstrating a good agreement with the analytical results. 
We thus provide further evidence that the phase space methods is a valuable tool for the studies of phase transitions 
and also open a new way to detect ESQPTs.

\end{abstract}

\date{\today}

\maketitle

\section{introduction}
 
 The pioneering works of Weyl \cite{Weyl1927} and Wigner \cite{Wigner1932} have triggered
 tremendous efforts to develop the so called phase space methods 
 \cite{Weyl1950,Zachos2005,Schroeck2013,Hillery1984,Lee1995,Polkovnikov2010}. 
 In this approach, a quantum state is described by a quasiprobability distribution defined 
 in the classical phase space instead of the density matrix in Hilbert space
 \cite{Wigner1932,Husimi1940,Glauber1963,Weinbub2018,Seyfarth2020,Koczor2020}. 
 Consequently, the expectations of quantum operators are reformulated as average of their classical counterpart over 
 the classical phase space in novel algebraic ways. 
 The quantum mechanics is, therefore, interpreted as a statistical theory 
 on the classical phase space \cite{Moyal1949,Takabayasi1954}. 
 This further leads to the phase space methods can provide 
 valuable insights into the correspondence between 
 quantum and classical systems \cite{Torres1990,Bohigas1993}. 
 As an alternative formulation of quantum mechanics, 
 phase space methods has numerous applications in many areas of physics, 
 including quantum optics \cite{Schleich2011}, atomic physics \cite{Mahmud2005,Blakie2008}, 
 quantum chaos \cite{Nonnenmacher1998,Toscano2008}, 
 condensed matter physics \cite{Aulbach2004,Carmesion2020},
 and quantum thermodynamics \cite{Altland2012a,Altland2012b,Brodier2020}.
 In particular, recent studies have been found that phase space methods acts as a powerful tool 
 for studying the quantum phase transitions in many-body systems 
 \cite{Romera2012,Calixto2012,Romera2014,Octavio2015,Calixto2015,Castanos2018,Mzaouali2019,Lopez2020}.
 
 In this work, we give further verifications of the usefulness of the 
 phase space methods in the studies of phase transitions. 
 To this end, we analyze the phase space signatures of the excited state quantum phase transitions (ESQPTs). 
 As a generalization of the ground state quantum phase, an ESQPT is
 characterized by the divergence in the density of states at the critical energy $E_c$ \cite{Caprio2008,Stransky2014}.
 Different kinds of ESQPTs have been identified, 
 both theoretically \cite{Brandes2013,Magnani2014,Bastidas2014,Stransky2016,Puebla2016,Rodriguez2018,ZhuG2019} 
 and experimentally \cite{Larese2013,Dietz2013,TianT2020}, in various many body systems.
 These works have derived much efforts to look at the effects of ESQPTs on the nonequilibrium properties 
 of quantum many body systems
 \cite{Relano2008,Perez2009,Perez2011,Engelhart2015,Santos2015,Santos2016,PerezS2017,Kloc2018,
 WangQ2019a,Cameo2020,WangQ2020}. 
 Such studies are in turn opened new avenues for detecting ESQPTs through the nonequilibrium 
 quantum dynamics in many body systems \cite{Puebla2013a,Puebla2013b,WangQ2017,WangQ2019b}, 
 which can be accessed within current experimental technologies \cite{Polkovnikov2011}.
 In addition, the relations between ESQPTs and the onset of chaos, the thermal phase transitions, as well as the
 exception points in non-Hermitian systems are also received a great deal of attention
 \cite{PerezF2011,Lobez2016,Perez2017,Sindelks2017}. 
 In spite of these developments, a complete understanding of ESQPTs is still lack.
 
 Here, from the phase space perspective, we focus on the phase space signatures of ESQPTs 
 in different many body systems.
 Specifically, we use the Husimi quasiprobability function and its associated marginal distributions 
 to analyze the ESQPTs in Lipkin and coupled top models, respectively.
 We first consider the dynamics of the Husimi function following a sudden quench process, and then focus on the
 properties of the long time averaged Husimi function and its marginal distributions.
 We find that the time evolution of the Husimi function undergoes a remarkable change 
 as the quench parameter passes through the critical point of an ESQPT.
 The presence of an ESQPT can be clearly identified by the particular behavior in the dynamics of the Husimi function.
 For the long time averaged Husimi function and its marginal distributions, we again find
 sharp signatures of the ESQPT in their properties.
 In addition, we discuss how to extract the critical points of the ESQPT using the second moment 
 and Wherl entropy of the long time averaged Husimi function, 
 showing a good agreement between the numerical and analytical results.
 Hence, our analysis places the phase space methods as a useful tool in the study of ESQPTs.

 The article is organized as follows.
 In Sec.~\ref{Secd}, we introduce the Husimi function and its marginal distributions, as well as the definitions of  
 their second moment and Wehrl entropy.
 In Sec.~\ref{Third}, we present, respectively, the Hamiltonians of the Lipkin and coupled top models 
 with briefly review their basic features, mainly focus on the properties of ESQPT.
 In Sec.~\ref{Four}, we report our results with respective to Lipkin and coupled top models. 
 We finally summarize and conclude our results in Sec.~\ref{Fv}.

\section{Husimi function} \label{Secd}

As the Gaussian smoothing of the Wigner function, the Husimi function, also known as $Q$ function, 
is a positive-definite function and defined as \cite{Husimi1940,Lee1995}
\be \label{Hfdf}
   Q(p,q)=\la\zeta(p,q)|\rho|\zeta(p,q)\ra,
\ee
where $\rho=|\psi\ra\la\psi|$ being a quantum state of the system and $|\zeta(p,q)\ra$
denotes the minimal uncertainty (coherent) state centered in the phase space point $(p,q)$ with
$p$ and $q$ are the canonical momentum and position, respectively. 
It is known that the coherent state covers a phase space region centered at $(p,q)$ with volume $\hbar$,
therefore, the Husimi function can be considered as the probability of observing 
the system in that region \cite{Furuya1992}.
In the following of our study, we set $\hbar=1$.

  \begin{figure}
    \includegraphics[width=\columnwidth]{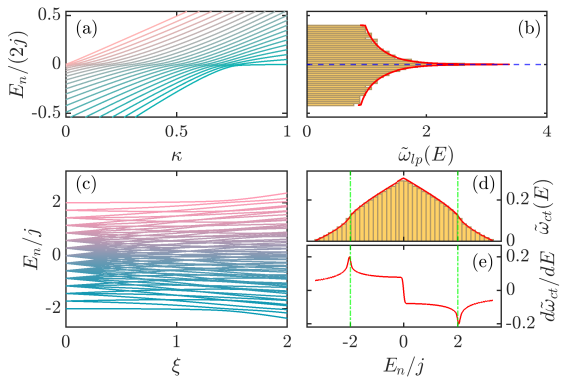}
  \caption{(a) Energy spectrum of the Lipkin model as a function of $\kappa$ with $j=N/2=20$.
  (b) Rescaled density of states, $\tilde{\omega}_{lp}(E)=\omega_{lp}(E)/j$, of the Lipkin model 
  with $\kappa=0.4$ and $j=N/2=500$.
  The red solid line is the semiclassical result and 
  the horizontal blue dashed line denotes the critical energy $E/(2j)=E_c/(2j)=0$.
  (c) Energy spectrum of the coupled top model as a function of $\xi$ with $j=7$.
  (d) Rescaled density of states, $\tilde{\omega}_{ct}(E)=\omega_{ct}(E)/j^2$, of the coupled top model 
  with $\xi=3$ and $j=70$.
  The red solid line denotes the semiclassical result.
  (e) Derivative of $\tilde{\omega}_{ct}(E)$ for the coupled top model with $\xi=3$ and $j=70$.
  Two vertical green dot-dashed lines in panels (d) and (e) indicate the critical energies $E/j=E_c/j=\pm2$ 
  of ESQPTs in coupled top model.}
  \label{DoSM}
 \end{figure} 

For spin systems that studied in this work, the Husimi function can be calculated by using the so-called 
$\mathrm{SU}(2)$ spin-$j$ coherent states \cite{Gilmore1990,Gazeau2009}
\be
  |\zeta\ra=(1+|\zeta|^2)^{-j}e^{\zeta J_+}|j,-j\ra,
\ee
where $\zeta\in\mathbb{C}, J_+=J_x+iJ_y$ is the spin raising operator, and $|j,-j\ra$ is the eigenstate of $J_z$ 
with eigenvalue $-j$, that is, $J_z|j,-j\ra=-j|j,-j\ra$. 
Here $J_{\{x,y,z\}}$ are the components of spin angular momentum operator.
The coherent states is an overcomplete set and satisfy the closure relation
\be \label{Csnrm}
   \frac{2j+1}{\pi}\int|\zeta\ra\la\zeta|\frac{d^2\zeta}{(1+|\zeta|^2)^2}=\mathbf{1},
\ee
with $d^2\zeta=d\mathrm{Re}(\zeta)d\mathrm{Im}(\zeta)$ being the integration measure on $\mathbb{C}$.
To visualize the Husimi function in phase space $(p,q)$, we parameterize $\zeta$ in terms of 
canonical variables $q$ and $p$ as \cite{Aguiar1992,Cameo2020}
\be
   \zeta(p,q)=\frac{q-ip}{\sqrt{4-(p^2+q^2)}},
\ee
with $p^2+q^2\leq4$.
Then it is straightforward to find that in phase space the closure relation Eq.~(\ref{Csnrm}) can be rewritten as
\be
   \frac{2j+1}{4\pi}\int_\Omega|\zeta(p,q)\ra\la\zeta(p,q)|dpdq=\mathbf{1},
\ee 
where the area $\Omega$ is defined by $p^2+q^2\leq4$.
Hence the normalization condition of the Husimi function Eq.~(\ref{Hfdf}) in phase space is given by
\be
   \frac{2j+1}{4\pi}\int_\Omega Q(p,q)dpdq=1.
\ee

As is well known, a great amount of information about the features of the system can be extracted from the 
moments of the Husimi function \cite{Aulbach2004,Romera2012,Calixto2012,Romera2014}. 
Among all moments, an important and useful one is the second moment (also dubbed as
the generalized inverse participation ratio), 
which quantifies the degree of delocalization of a quantum state in phase space, and read as
\be \label{SdM}
   M_2=\frac{2j+1}{4\pi}\int_\Omega Q^2(p,q)dpdq.
\ee
For an extremely extended state, the phase space is uniformly covered by state $\rho=|\psi\ra\la\psi|$,
we would have $Q(p,q)\sim1/(2j+1)$. In this case, one can find that $M_2\sim1/(2j+1)$ which goes to zero as $j\to\infty$.
Hence, the smaller is the value of $M_2$, the higher is the degree of delocalization of state $|\psi\ra$ in phase space.
On the other hand, if the state $\rho=|\psi\ra\la\psi|$ is identical to one point $(p_0,q_0)$ in phase space, we would expect
$Q(p,q)\sim\exp{\left[-(q-q_0)^2/(2\sigma_q^2)+(p-p_0)^2/(2\sigma_p^2)\right]}$ with, according to normalization condition, 
$\sigma_q\sigma_p=2/(2j+1)$.
In the classical limit $j\to\infty$, one can see that $Q(p,q)$ shrinks to the point $(p_0,q_0)$, as expected.
Then, the second moment of Husimi function for this maximal localized state is given by $M_2\sim1/2$.
Therefore, we have $M_2\in[0,1/2]$ 
with the maximum value corresponds to the maximum localization state in phase space.

Besides the second moment of the Husimi function, another quantity that has been employed in various studies 
to characterize the properties of the Husimi function is the Wehrl entropy 
\cite{Romera2012,Calixto2012,Romera2014,Octavio2015,Calixto2015}.
As the classical counterpart of the quantum von Neumann entropy, the Wehrl entropy is defined as \cite{Wehrl1979}
\be \label{Whp}
   W=-\frac{2j+1}{4\pi}\int_\Omega Q(p,q)\ln[Q(p,q)]dpdq.
\ee
Here, it is worth pointing out that the second moment $M_2$ in Eq.~(\ref{SdM}) 
can be considered as a linearized version of the Wehrl entropy.
Therefore, the Wehrl entropy also measures the degree of localization of a quantum state in phase space.
However, in contrast to $M_2$, the degree of delocalization increases with increasing $W$.
For the fully extended states, we have $W_{max}\sim\ln(2j+1)$.
In addition, the Lieb conjecture shows that the minimum Wehrl entropy is
$W_{min}=j/(j+1)$, so that $W_{min}\to1$ as $j\to\infty$ \cite{Lieb2002}.  

  \begin{figure*}
    \includegraphics[width=\textwidth]{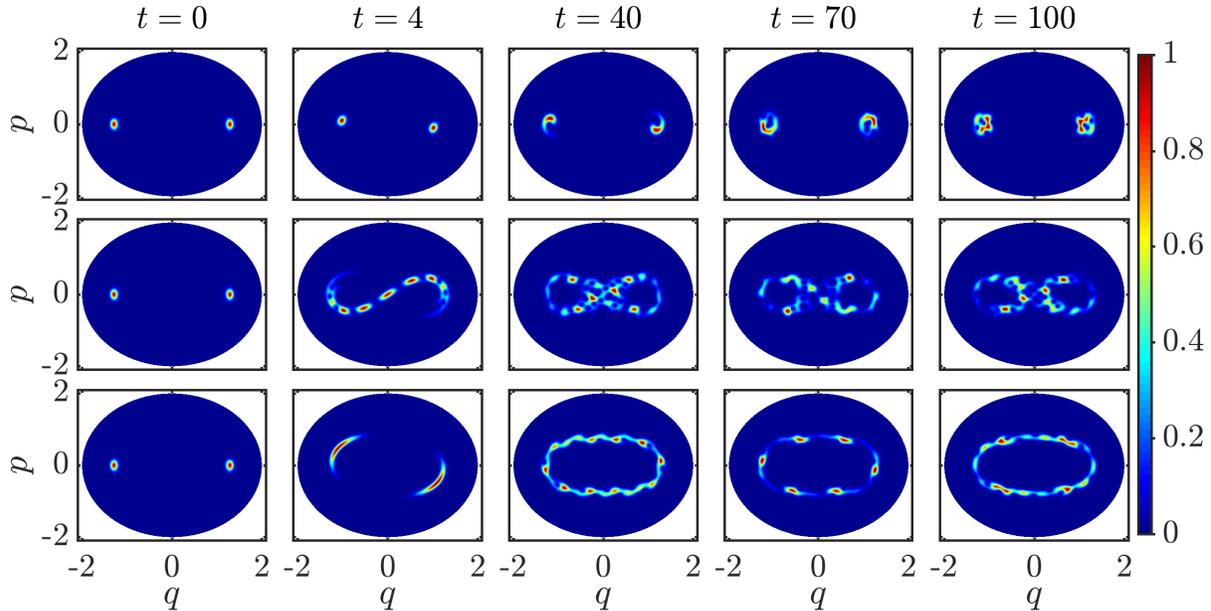}
  \caption{Snapshots of the rescaled spin Husimi function $\mathcal{Q}_t(p,q)=Q_t(p,q)/Q_t^m$ with $Q_t^m$ 
  being the maximum value of $Q_t(p,q)$, 
  at different time steps for the Lipkin model with $\eta=0.4,1,1.7$ (from top to bottom).
  Other parameters are: $j=N/2=200$ and $\kappa=0.4$.
  }
  \label{timeHF}
 \end{figure*}

More insights into the phase space features of a state can be obtained from the marginal distributions 
of the Husimi function for position and momentum spaces, respectively.
For spin-$j$ coherent states studied in this work, they are defined as
\begin{align} \label{MgHf}
   &Q(q)=\sqrt{\frac{2j+1}{4\pi}}\int Q(p,q)dp, \notag \\
   &Q(p)=\sqrt{\frac{2j+1}{4\pi}}\int Q(p,q)dq,
\end{align} 
with normalization conditions
\be
   \sqrt{\frac{2j+1}{4\pi}}\int Q(q)dq=\sqrt{\frac{2j+1}{4\pi}}\int Q(p)dp=1.
\ee
Accordingly, the second moment and Wehrl entropy of marginal distributions are given by
\begin{align} \label{SWmg}
   &M_2^{(\mu)}=\sqrt{\frac{2j+1}{4\pi}}\int Q^2(\mu)d\mu, \notag \\
   &W^{(\mu)}=-\sqrt{\frac{2j+1}{4\pi}}\int Q(\mu)\ln[Q(\mu)]d\mu,
\end{align}
where $\mu\in\{p,q\}$.

We would like to point out that the marginal distributions $Q(q), Q(p)$ of the Husimi function 
do not equal to the density functions
$|\la q|\psi\ra|^2$ and $|\la p|\psi\ra|^2$, in sharp contrast to the case of Wigner function. 
In fact, they are the Gaussian smeared density distribution in position and 
momentum spaces, respectively \cite{Romera2012,Varga2003}.
Note further that, in general, we have $|M_2-M_2^{(p)}M_2^{(q)}|=\delta M_2\geq0$ 
and $|W-[W^{(p)}+W^{(q)}]|=\delta W\geq0$
with $\delta M_2, \delta W$ decrease as the system size increases except around some singular points, such as 
quantum critical points \cite{Aulbach2004,Romera2012,Varga2003}. 

In the following, by exploiting above outlined properties of the Husimi function, we will explore the 
signatures of ESQPTs in two many body systems, namely the Lipkin and coupled top models.

  \begin{figure}
  \includegraphics[width=\columnwidth]{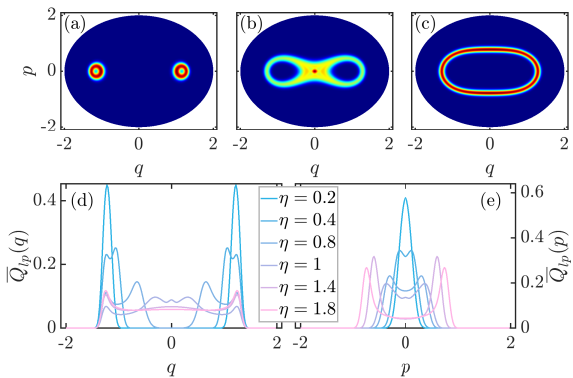}
  \caption{Rescaled long-time averaged Husimi function 
  $\overline{\mathcal{Q}}_{lp}(p,q)=\overline{Q}_{lp}(p,q)/\overline{Q}_{lp,m}$
  of the Lipkin model for (a) $\eta=0.4$, (b) $\eta=1$, and (c) $\eta=1.8$ with $j=N/2=200$ and $\kappa=0.4$.
  Here $\overline{Q}_{lp,m}$ denotes the maximum value of $\overline{Q}_{lp}(p,q)$.
  The color scale in Fig.~\ref{timeHF} has been used for panels (a)-(c).
  Marginal distributions $\overline{Q}_{lp}(q)$ and $\overline{Q}_{lp}(p)$ of $\overline{Q}_{lp}(p,q)$ 
  for several values of $\eta$ are plotted in panels
  (d) and (e), respectively.}
  \label{AvgHf}
 \end{figure}

\section{Models} \label{Third}

\subsection{Lipkin model}

 The Lipkin model describes $N$ spin-$1/2$ particles with infinite range of interactions and subjected to an
 external magnetic field. 
 It was first introduced as a toy model to explore phase transitions in nuclear systems \cite{Lipkin1965} and
 since then it has been exploited as a paradigmatic model in 
 various studies of quantum phase transitions 
 \cite{Ribeiro2008,Engelhardt2013,Botet1983,Dusuel2005,Castanos2006,Campbell2016}.   
 As the Lipkin model has broad applications in several different fields of physics, 
 it has attracted much attention from both theoretical 
 \cite{BaoJ2020,Lourenco2020,Russomanno2017,HuangY2018}
 and experimental \cite{Morrison2008,Zibold2010}
 perspectives in recent years.
 
 By using the collective operators $J_\alpha=\sum_k^N\sigma_k^\alpha/2, \{\alpha=x,y,z\}$ with $\sigma_k^\alpha$
 is the $\alpha$th component of the Pauli matrix of $k$th spin,        
 the Hamiltonian of the Lipkin model can be written as 
 \be \label{LMG}
  H_{lp}=-\frac{4(1-\kappa)}{N}J^2_{x}+\kappa\left(J_z+\frac{N}{2}\right),
 \ee
 where $\kappa$ denotes the strength of the external magnetic field.
 The total spin operator $\mathbf{J}^2=\mathbf{J}_x^2+\mathbf{J}_y^2+\mathbf{J}_z^2$ with eigenvalue $j(j+1)$
 is commutated with the Hamiltonian. 
 In our study, we restrict ourself in the spin sector with $j=N/2$, thus
 the dimension of the Hamiltonian matrix is $\mathcal{D}_{H_{lp}}=N+1$.
 Moreover, due to conservation of the parity operator $\hat{\Pi}_{lp}=e^{i\pi(j-m)}$ with $m\in\{-j,-j+1,\ldots,j\}$ 
 is the eigenvalue of $J_z$, 
 the Hamiltonian matrix is further split into even- and odd- parity blocks 
 with dimensions $\mathcal{D}_{H_{lp}}^e=N/2+1$ and
 $\mathcal{D}_{H_{lp}}^o=N/2$, respectively. 
 We focus on the even-parity block which includes the ground state of the system.
 
 The Lipkin model undergoes a second-order ground state quantum phase transition 
 from the paramagnetic phase with $\kappa<4/5$
 to the ferromagnetic phase with $\kappa>4/5$ at the critical point $\kappa_c=4/5$ 
 \cite{Botet1983,Dusuel2005,Castanos2006}.
 The ground state quantum phase transition of 
 the Lipkin model has been studied extensively in numerous works 
 \cite{Botet1983,Dusuel2005,Castanos2006,Campbell2016,BaoJ2020,Lourenco2020,Latorre2005,Titum2020}.
 In particular, the phase space
 characters of the ground state quantum phase transition of the Lipkin model 
 has been explored in Ref.~\cite{Romera2014}. 
 Here, we are interested in analyzing the signatures of ESQPT in phase space 
 by means of the Husimi function.
 The Lipkin model exhibits an ESQPT at critical energy $E_c=0$ when $\kappa<4/5$ \cite{Caprio2008,Perez2009}.
 The ESQPT in Lipkin model is characterized by
 the singular behavior in its density of states
 $\omega_{lp}(E)=\sum_n\delta(E-E_n)$. 
 
 In Fig.~\ref{DoSM}(a), we plot the energy levels of 
 the Lipkin model with $j=N/2=20$ as a function of $\kappa$. 
 We can see that the energy levels exhibit an obvious collapse around $E_c=0$ for the cases of $\kappa<4/5$.
 This means the density of states of the Lipkin model would have a sharp peak 
 in the neighborhood of $E_c$. 
 Indeed, as can be seen from Fig.~\ref{DoSM}(b), both the numerical and semiclassical results \cite{Perez2009} 
 show that, at the critical energy $E_c=0$,
 $\omega_{lp}(E)$ has a cusp singular which 
 turns into a logarithmic divergence as $j=N/2\to\infty$ \cite{Ribeiro2008,Stransky2014}.

 \subsection{Coupled top model}
 
 The second model we considered is the so-called coupled top model, also known as the Feingold-Peres model 
 \cite{Feingold1983,Feingold1984,Hines2005,FanY2017,Mondal2020}.
 It describes the interaction between two larger spins with respective angular momentum operators 
 $\mathbf{J}_1=(J_{1x},J_{1y},J_{1z})$ and $\mathbf{J}_2=(J_{2x},J_{2y},J_{2z})$, 
 whose Hamiltonian takes the form
 \be \label{CTPH}
    H_{ct}=J_{1z}+J_{2z}+\frac{\xi}{j}J_{1x}J_{2x},
 \ee
 where $\xi$ is the coupling strength between two spins.
 Here, we assume two spins have same magnitude $\mathbf{J}_1^2=\mathbf{J}_2^2=j(j+1)$,
 so that the dimension of the Hilbert space is $\mathcal{D}_{H_{ct}}=(2j+1)^2$. 
 However, as the Hamiltonian in Eq.~(\ref{CTPH}) remains invariant under the permutation 
 symmetry $\hat{\mathscr{P}}$ between two spins $(J_1\leftrightarrow J_2)$ 
 and under parity $\hat{\Pi}_{ct}=e^{i\pi(2j-m_1-m_2)}$
 with $m_1,m_2\in\{-j,-j+1,\ldots,j\}$ are the eigenvalues of $J_{1z}, J_{2z}$, the Hilbert space can be 
 further decomposed into four subspaces according to the eigenvalues of $\hat{\mathscr{P}}$ and $\hat{\Pi}_{ct}$.
 We shall focus on the subspace identified by $\mathscr{P}=+1, \Pi_{ct}=+1$, denoted by $V_{++}$, 
 which contains the ground state.
 We also restrict to integer $j$, thus the dimension of $V_{++}$ is $\mathcal{D}_{V_{++}}=(j+1)^2$ \cite{FanY2017}.

  \begin{figure}
  \includegraphics[width=\columnwidth]{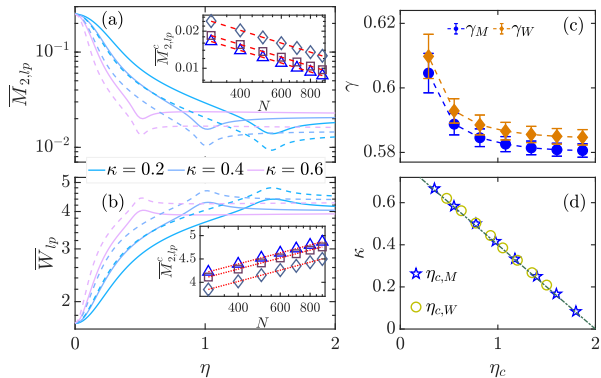}
  \caption{(a) Second moment of $\overline{Q}_{lp}(p,q)$ as a function of $\eta$ for several $\kappa$ 
  with $j=N/2=200$ (solid lines) and $j=N/2=400$ (dashed lines).
  Inset: Critical second moment, $\overline{M}_{2,lp}^c=\overline{M}_{2,lp}(\eta_c)$, as a function of $N$
  for $\eta_c=1.5$ (triangles), $\eta_c=1$ (squares), and $\eta_c=0.5$ (diamonds).
  Red dashed lines are of the form $\overline{M}_{2,lp}^c\sim N^{-\gamma_M}$.
  (b) Wehrl entropy of $\overline{Q}_{lp}(p,q)$ as a function of $\eta$ for various $\kappa$ 
  with $j=N/2=200$ (solid lines) and $j=N/2=400$ (dashed lines).
  Inset: Critical Wehrl entropy, $\overline{W}_{lp}^c=\overline{W}_{lp}(\eta_c)$, as a function of $N$ for 
  $\eta_c=1.5$ (triangles), $\eta_c=1$ (squares), and $\eta_c=0.5$ (diamonds).
  Red dotted lines are of the form $\overline{W}_{lp}^c\sim \gamma_W\ln(N)$.
  (c) Values of the finite size scaling exponents of $\overline{M}_{2,lp}^c$ and $\overline{W}_{lp}^c$
  for several $\eta_c$ with $j=N/2=200$. 
  (d) Critical values $\eta_{c,M}, \eta_{c,W}$ estimated from the extrema of $\overline{M}_{2,lp}$ and $\overline{W}_{lp}$,
  respectively, for different values of $\kappa$ with $j=N/2=400$. The dot-dashed line indicates the 
  analytical result in Eq.~(\ref{Crp}).}
  \label{smwp}
 \end{figure}

 The coupled top model has been studied extensively in diverse fields of physics
 \cite{Hines2005,FanY2017,Mondal2020,Robb1998,Sayak2019}.
 It is known that its ground state displays a second-order quantum phase transition at $\xi_c=1$, 
 which separates the ferromagnetic phase 
 with $\xi<\xi_c$ from the paramagnetic phase with $\xi>\xi_c$ \cite{Hines2005,Mondal2020}. 
 In particular, it has been found that the coupled top model undergoes ESQPTs at critical energies 
 $E_c/j=\pm2$ for $\xi>\xi_c=1$.
 Different from the case of Lipkin model, the ESQPTs in the coupled top model are identified by the non-analytical 
 behaviors in the derivative of the density of states at the critical energies. 
 A very similar signature of ESQPT has also been 
 founded in the Dicke model \cite{Brandes2013,Magnani2014}. 
 
  \begin{figure}
  \includegraphics[width=\columnwidth]{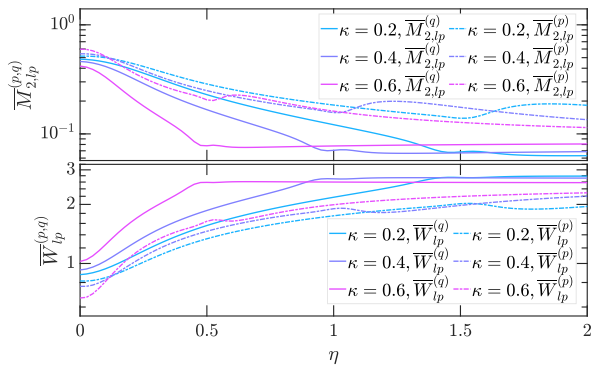}
  \caption{Second moment (upper panel) and Wehrl entropy (bottom panel) of 
  the marginal distributions of $\overline{Q}_{lp}(p,q)$ as a function of $\eta$ 
  for several $\kappa$ with $j=N/2=200$.}
  \label{SMWmgd}
 \end{figure}

 The energy spectrum of the coupled top model as a function of control parameter $\xi$ is plotted 
 in panel (c) of Fig.~\ref{DoSM} for $j=7$. 
 We see that the energy spectrum becomes more complex as the value of $\xi$ increases. 
 However, the collapse of the energy levels around the critical energy observed 
 in the Lipkin model [cf.~Fig.~\ref{DoSM}(a)] 
 does not exist in the energy spectrum of the coupled top model.
 This means the density of states of the coupled top model, denoted by $\omega_{ct}(E)$, 
 will behave as a continuous function of energy at $\xi>\xi_c$, as is shown in Fig.~\ref{DoSM}(d) 
 for the numerical and associated semiclassical results \cite{Mondal2020}.
 In fact, the ESQPTs in the coupled top model are uncovered through the 
 singular behaviors in the derivative of $\omega_{ct}(E)$.
 In panel (e) of Fig.~\ref{DoSM}, we plot the derivative of $\tilde{\omega}_{ct}(E)=\omega_{ct}(E)/j^2$ 
 as a function of $E_n/j$ with $j=70$.
 As it can be seen, $d\tilde{\omega}_{cp}(E)/dE$ develops a cusp singular at the critical energies. 
 Such singularities are expect to be the logarithmic divergences in the thermodynamic limit $j\to\infty$ 
 \cite{Stransky2014,Stransky2016}. 
 In the following, we will constraint ourselves to the critical energy $E_c/j=-2$, 
 since $E_c/j=2$ gives rise the same results.  
 
  \begin{figure*}
    \includegraphics[width=\textwidth]{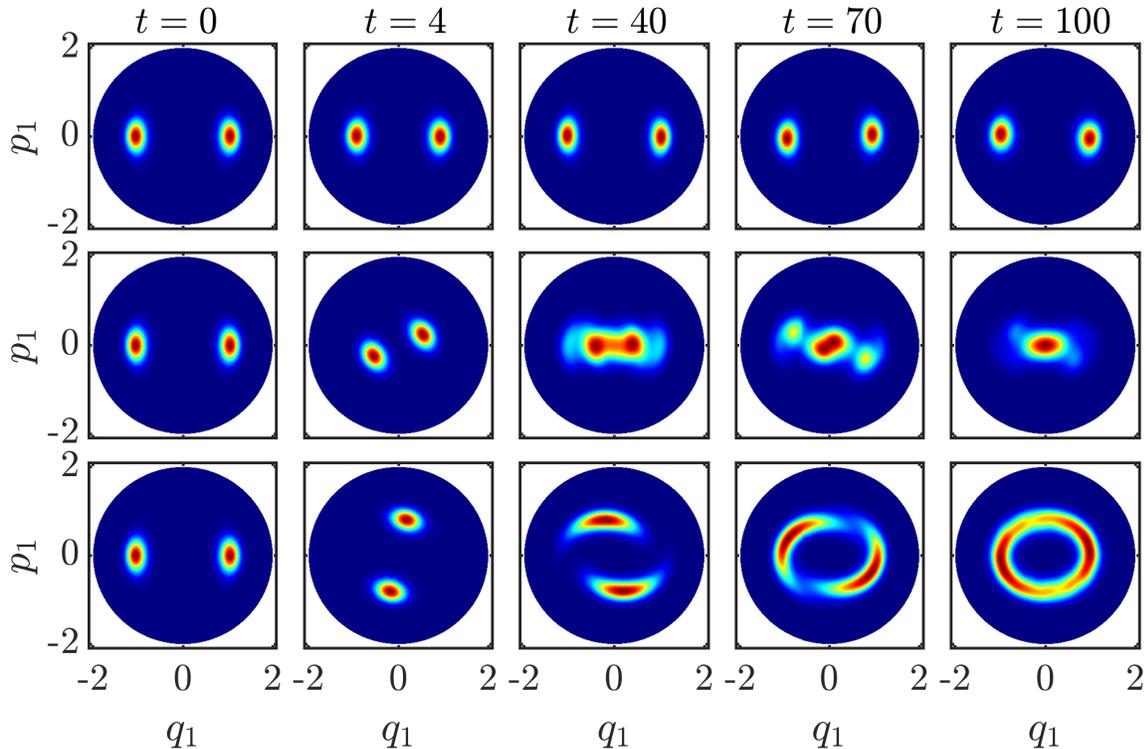}
  \caption{Snapshots of the rescaled spin Husimi function $\mathcal{Q}_t(p_1,q_1)=Q_t(p_1,q_1)/Q_t^m(p_1,q_1)$ 
  with $Q_t^m(p_1,q_1)$ being the maximum value of $Q_t(p_1,q_1)$, 
  at different time steps for the coupled top model with $\xi_1=2.5,1.5,0.5$ (from top to bottom).
  Other parameters are: $j=30$ and $\xi_0=3$. The color scale of figure \ref{timeHF} has been used.
  }
  \label{timeHFCT}
 \end{figure*}

\section{Results and discussions} \label{Four}

 In this section, we discuss how to identify the signatures of ESQPT from the perspective of 
 quantum phase space by means of the Husimi function in two aforementioned models. 
 We consider the the impacts of ESQPT on the dynamical features of Husimi function 
 using the quantum quench protocol and
 focus on the properties of the long-time averaged Husimi function.
 
 \subsection{Husimi function of the Lipkin model}
 
 For the Lipkin model, the quantum quench protocol is described as follows.
 The model is initially prepared in the ground state $|\psi_0\ra$ of $H_{lp}$ with $0<\kappa<4/5$.
 At $t=0^+$, we suddenly add an external magnetic field along $z$ direction with strength $\eta$,
 and let the model evolve under the Hamiltonian $H_{lp}^f=H_{lp}+\eta(S_z+N/2)$. 
 For a certain value of $\kappa$, one can take the model crossing of the critical energy of ESQPT 
 by varying the strength of $\eta$. 
 The critical strength $\eta_c$, which leads to the critical energy $E_c=0$, can be obtained through 
 the coherent state approach and is given by \cite{,Relano2008,Perez2009}
 \be \label{Crp}
    \eta_c=2-\frac{5}{2}\kappa,
 \ee
 with $0<\kappa<4/5$. 
 We stress that the critical strength $\eta_c$ for the ESQPT is smaller than 
 the quench strength which drive the model through the ground state quantum phase transition \cite{Perez2009}.

 The quantum state of the model is evolved as 
 $\rho(t)=|\psi(t)\ra\la\psi(t)|=e^{-iH_{lp}^ft}\rho(0)e^{iH_{lp}^ft}$ with $\rho(0)=|\psi_0\ra\la\psi_0|$.
 Hence the Husimi function at time $t$ can be written as
 \begin{align}
    Q_t(p,q)&=\la\zeta(p,q)|\rho(t)|\zeta(p,q)\ra \notag \\
          &=\left|\sum_ne^{-iE_nt}\la\zeta(p,q)|E_n\ra\la E_n|\psi_0\ra\right|^2,
 \end{align}
 where $|E_n\ra$ is the $n$th eigenstate of $H_{lp}^f$ with eigenvalue $E_n$. 
 Here we see that $Q_t(p,q)$ is strongly depended on the transition amplitudes between the initial state and the $n$th
 eigenstate of $H_{lp}^f$.
 
  \begin{figure}
  \includegraphics[width=\columnwidth]{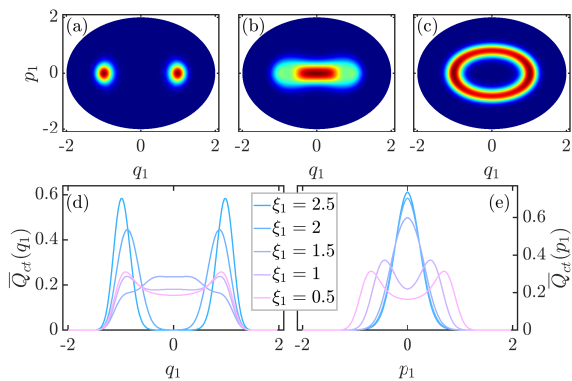}
  \caption{Rescaled long-time averaged Husimi function 
  $\overline{\mathcal{Q}}_{ct}(p_1,q_1)=\overline{Q}_{ct}(p_1,q_1)/\overline{Q}_{ct,m}$
  of the coupled top model for (a) $\xi_1=2.5$, (b) $\xi_1=1.5$, and (c) $\xi_1=0.5$ with $j=30$ and $\xi_0=3$.
  Here $\overline{Q}_{ct,m}$ denotes the maximum value of $\overline{Q}_{ct}(p_1,q_1)$.
  The color scale in Fig.~\ref{timeHF} has been employed for panels (a)-(c).
  Marginal distributions $\overline{Q}_{ct}(q_1)$ and $\overline{Q}_{ct}(p_1)$ of $\overline{Q}_{ct}(p_1,q_1)$ 
  for several values of $\xi_1$ are plotted in panels
  (d) and (e), respectively.}
  \label{AvgctHf}
 \end{figure}

 In Fig.~\ref{timeHF}, we plot the Husimi function of the Lipkin model at different time steps for several values of 
 $\eta$ with $\kappa=0.4$ and $N=400$. 
 For this case, the critical quench strength in Eq.~(\ref{Crp}) is $\eta_c=1$.
 We first note that the ground state in the even-parity sector can be well described by the so-called even 
 coherent states \cite{Castanos2006,Romera2014,Dodonov1974}, 
 $|\zeta(p,q)\ra_+=\mathcal{N}_+(p,q)[|\zeta(p,q)\ra+|\zeta(-p,-q)\ra]$,
 with $\mathcal{N}_+(p,q)=1/\sqrt{2\{1+[1-(p^2+q^2)/2]^{2j}\}}$ is the normalization constant. 
 As a result, the Husimi function of initial state should be represented by two 
 symmetrically localized packets in phase space, as seen in the first column of Fig.~\ref{timeHF}. 
 As time increases, the Husimi function exhibits remarkable different behaviors for $\eta$ below, at, 
 and above the critical value $\eta_c=1$.
 Specifically, as observed in the top row of Fig.~\ref{timeHF},
 the Husimi function remains as two distinct localized packets in its time evolution for $\eta<\eta_c$.
 At the critical point $\eta=\eta_c$ [see the second row in Fig.~\ref{timeHF}], 
 the evolution of the Husimi function results in an extension in phase space 
 and, in particular, two initially disconnected packets are joined together in this case. 
 Finally, when $\eta>\eta_c$, the two initially separated packets are merged into a single one and 
 the evolution of the Husimi function is in sharp contrast to the case of $\eta<\eta_c$, 
 as illustrated in the last row of Fig.~\ref{timeHF}. 
 The strikingly distinct behaviors in the dynamics of Husimi function 
 on two sides of the transition suggest that the underlying ESQPT has non-trivial impacts on the dynamics of the model.
 Moreover, the particular dynamical behavior of Husimi function at $\eta=\eta_c$ can be employed to probe the 
 occurrence of an ESQPT.

 To get more evident signatures of ESQPT, we consider the long-time averaged Husimi function
 \be \label{AvHf}
     \overline{Q}(p,q)=\la\zeta(p,q)|\bar{\rho}|\zeta(p,q)\ra,
 \ee
 where $\bar{\rho}$ is the long-time averaged state of the model and defined as
 \be
    \bar{\rho}=\lim_{T\to\infty}\frac{1}{T}\int_0^T\rho(t)dt.
 \ee
 For the Lipkin model, it is straightforward to find that the explicit expression 
 of the long-time averaged Husimi function is given by
 \be
    \overline{Q}_{lp}(p,q)=\sum_n|\la\zeta(p,q)|E_n\ra|^2|\la E_n|\psi_0\ra|^2.
 \ee
 Here we see again the transition probabilities between $|\psi_0\ra$ and the $n$th eigensate of $H_{lp}^f$
 play crucial role in determining the behaviors of $\ov{Q}_{lp}(p,q)$.
 
 In Figs.~\ref{AvgHf}(a)-\ref{AvgHf}(c), we plot $\overline{Q}_{lp}(p,q)$ for various values of $\eta$.
 We see that the structure of $\overline{Q}_{lp}(p,q)$ changes drastically as $\eta$ 
 passes through the critical point. 
 For $\eta<\eta_c$, $\overline{Q}_{lp}(p,q)$ consists of two localized and disconnected parts.
 With increasing $\eta$, the extension of $\overline{Q}_{lp}(p,q)$ leads to two disconnected parts
 joining together at $\eta=\eta_c=1$. 
 As $\eta$ increases further, the two joined parts are merged into a single one.
 We also note that $\overline{Q}_{lp}(p,q)$ has a rather larger degree of delocalization at $\eta=\eta_c$.
 The features of $\overline{Q}_{lp}(p,q)$ are more visible in its
 marginal distributions [cf.~Eq.(\ref{MgHf})]. 
 For several values of $\eta$, the marginal distributions $\overline{Q}_{lp}(q)$ and $\overline{Q}_{lp}(p)$ are plotted 
 in Figs.~\ref{AvgHf}(d) and \ref{AvgHf}(e), respectively.
 Clearly, the width of $\overline{Q}_{lp}(q)$ and $\overline{Q}_{lp}(p)$ increase with increasing $\eta$ due to
 the extension of $\overline{Q}_{lp}(p,q)$ in phase space.
 Moreover, an obvious complex shape in the marginal distributions at $\eta_c=1$ implies that 
 they act as indicators of the ESQPT, in particular for the case of $\overline{Q}_{lp}(q)$. 
 
 To further elucidate the signatures of ESQPT in the properties of $\overline{Q}_{lp}(p,q)$,
 we evaluate its second moment $\overline{M}_{2,lp}$ [see Eq.~(\ref{SdM})] 
 and Wehrl entropy $\overline{W}_{lp}$ [see Eqs.~(\ref{Whp})].
 In Figs.~\ref{smwp}(a) and \ref{smwp}(b), we plot $\overline{M}_{2,lp}$ and $\overline{W}_{lp}$ as a function of $\eta$
 for several values of $\kappa$.
 We see that both the second moment and Wehrl entropy reach their extremum value at the critical value $\eta_c$.
 This means that the underlying ESQPT gives rise to a maximal extension of the quantum state.
 Moreover, the extremum values in the second moment and Wehrl entropy,
 denoted by $\overline{M}_{2,lp}^c$ and $\overline{W}^c_{lp}$, increase with increasing the system size $N$.
 In the insets of Fig.~\ref{smwp}(a) and \ref{smwp}(b), we show how $\overline{M}^c_{2,lp}$ 
 and $\overline{W}^c_{lp}$ vary with $N$ for several values of $\eta_c$.
 We find that $\overline{M}_{2,lp}^c$ follows 
 a power law scaling $\overline{M}_{2,lp}^c\sim N^{-\gamma_M}$, regardless of the value of $\eta_c$. 
 However, in all cases, $\overline{W}_{lp}^c$ exhibits a logarithmic scaling of the form 
 $\overline{W}_{lp}^c\sim\gamma_W\ln(N)$.
 The values of the scaling exponents $\gamma_M$ and $\gamma_W$ are demonstrated in Fig.~\ref{smwp}(c).
 As it can be seen, $\gamma_M$ and $\gamma_W$ decrease with an increases in $\eta_c$.
 By identifying the position of the extremum in $\overline{M}_{2,lp}$ and $\overline{W}_{lp}$ 
 as the estimation of the critical point, we compare the numerically obtained critical points
 with the analytical ones in Eq.~(\ref{Crp}).
 A good agreement between them can be clearly observed in Fig.~\ref{smwp}(d).  
 These results suggest that the second moment and Wehrl entropy of $\overline{Q}_{lp}(p,q)$ 
 can reliably detect ESQPTs in Lipkin model.
 
 Figure \ref{SMWmgd} displays the variation of the second moment and Wehrl entropy 
 of the marginal distributions with $\eta$ for several values $\kappa$.
 The underlying ESQPT induces the remarkable changes in the behaviors 
 of the marginal quantities, as is evident from Fig.~\ref{SMWmgd}.
 We further note that the extension of the quantum state in position direction 
 is larger than that in momentum direction consistent with the behaviors of 
 the marginal distributions observed in Figs.~\ref{AvgHf}(d) and \ref{AvgHf}(e).

  \begin{figure}
  \includegraphics[width=\columnwidth]{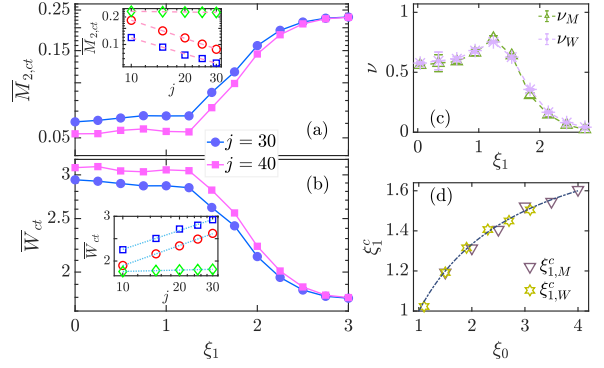}
  \caption{(a) Second moment of $\overline{Q}_{ct}(p_1,q_1)$ as a function of $\xi_1$ for different system size $j$ 
  with $\xi_0=3$ and $\xi_1^c=1.5$ [see Eq.~(\ref{Crxi})].
  Inset: $\overline{M}_{2,ct}$ as a function of $j$
  at $\xi_1=0.3$ (blue squares), $\xi_1=1.5$ (red circles), and $\xi_1=2.7$ (green diamonds).
  The dashed lines are of the form $\overline{M}_{2,ct}\sim j^{-\nu_M}$.
  (b) Wehrl entropy of $\overline{Q}_{ct}(p_1,q_1)$ as a function of $\xi_1$ for different $j$ 
  with $\xi_0=3$.
  Inset: $\overline{W}_{ct}$ as a function of $j$ for 
  $\xi_1=0.3$ (blue squares), $\xi_1=1.5$ (red circles), and $\xi_1=2.7$ (green diamonds).
  The dotted lines are of the form $\overline{W}_{ct}\sim \nu_W\ln(j)$.
  (c) Finite size scaling exponents $\nu_M$ and $\nu_W$
  as a function of $\xi_1$ with $\xi_0=3$. 
  (d) Estimated critical points $\xi_{1,M}^c, \xi_{1,W}^c$, obtained from the minima in $d\nu_{M(W)}/d\xi_1$,  
  as a function of $\xi_0$. The dot-dashed line denotes the 
  analytical result in Eq.~(\ref{Crxi}).}
  \label{MWct}
 \end{figure}

 \subsection{Husimi function of the coupled top model}
 
 To analyze the ESQPT in the coupled top model, the quench protocol consists as follows.
 Initially, the ground state $|\Psi_0\ra$ of $H_{ct}$ with $\xi=\xi_0>1$ is prepared.
 Then we suddenly change the coupling strength from $\xi_0$ to $\xi_1$ and consider 
 the evolution of the model governed by the Hamiltonian $H_{ct}(\xi_1)$.
 The critical coupling strength, denoted by $\xi_1^c$, is identified as the coupling that takes
 the energy of $H_{ct}(\xi_1)$ to the critical energy of the ESQPT, 
 so that $\la\psi(\xi_0)|H_{ct}(\xi_1^c)|\psi(\xi_0)\ra/j=E_c/j=-2$.
 By using the semiclassical approach (see Appendix \ref{AppA}), one can find $\xi_1^c$ is given by
 \be \label{Crxi}
   \xi_1^c=\frac{2\xi_0}{\xi_0+1},
 \ee
 with $\xi_0>1$.
 The critical coupling of the ESQPT depends on the value of $\xi_0$ and is always larger than the 
 ground state critical point $\xi_c=1$.  
 
 The evolved state of the model is 
 $\rho_t(\xi_1)=|\Psi_t\ra\la\Psi_t|=e^{-iH(\xi_1)t}\rho_0(\xi_0)e^{iH(\xi_1)t}$ 
 with $\rho_0(\xi_0)=|\Psi_0\ra\la\Psi_0|$.
 As the phase space of the coupled top model has $4$ dimensions,
 the evolved Husimi function expressed in terms of $\rho_t(\xi_1)$ takes the form
 \be
    Q_t(\bp,\bq)=\la\Upsilon(\bp,\bq)|\rho_t(\xi_1)|\Upsilon(\bp,\bq)\ra,
 \ee
 where $\bp=(p_1,p_2)$, $\bq=(q_1,q_2)$, and $|\Upsilon(\bp,\bq)\ra=|\zeta(p_1,q_1)\ra\otimes|\zeta(p_2,q_2)\ra$
 with
 \be
   \zeta(p_k,q_k)=\frac{q_k-ip_k}{\sqrt{4-(p_k^2+q_k^2)}},
   \qquad k=1,2.  \notag
 \ee
 The normalization condition for $Q_t(\bp,\bq)$ reads
 \be
   \left(\frac{2j+1}{4\pi}\right)^2\int_{\Omega_1}\int_{\Omega_2}Q_t(\bp,\bq)d\bp d\bq=1,
 \ee
 where $\Omega_1\in\{(p_1,q_1)|p_1^2+q_1^2\leq4\}$ and $\Omega_2\in\{(p_2,q_2)|p_2^2+q_2^2\leq 4\}$. 
 
 The four-dimensional Husimi function $Q_t(\bp,\bq)$ is difficult to display. 
 Therefore, we consider the projection of the Husimi function over the space $(p_1,q_1)$, so that
 $Q_t(p_1,q_1)\sim\int dp_2dq_2 Q_t(\bp,\bq)$.
 As the coherent states $|\zeta(p_2,q_2)\ra$ in the space $(p_2,q_2)$ 
 fulfill the normalization condition [cf.~Eq.~(\ref{Csnrm})], 
 the projected Husimi function adopts the form
 \be
    Q_t(p_1,q_1)=\la\zeta(p_1,q_1)|\rho_1^t(\xi_1)|\zeta(p_1,q_1)\ra,
 \ee
 with normalization condition 
 \be
   \frac{2j+1}{4\pi}\int_{\Omega_1}Q_t(p_1,q_1)dp_1dq_1=1. \notag
 \ee
 Here $\rho_1^t(\xi_1)=\mathrm{Tr}_2[\rho_t(\xi_1)]$ is the reduced density matrix of the first spin.
 
 In Fig.~\ref{timeHFCT}, we show the snapshots of the evolution of Husimi function at several time steps for
 different values of $\xi_1$ with $\xi_0=3$. 
 The critical value of $\xi_1$ for $\xi_0=3$ is $\xi_1^c=1.5$ [cf.~Eq.~(\ref{Crxi})].
 The ground state of the coupled top model has even-parity 
 and it can be well approximated by the even coherent states.
 The Husimi function at the initial time should consist of two distinct wave packets, which are symmetrically placed in 
 the phase space, as seen in the first column of Fig.~\ref{timeHFCT}.
 As time increases, the Husimi function of the coupled top model exhibits a very similar 
 behaviors as observed in the Lipkin model [cf.~Fig.~\ref{timeHF}].
 Namely, the evolution of the Husimi function remains as two different wave packets until $\xi_1=\xi_1^c$, 
 where two separated wave packets are joined together.  
 Further decreases $\xi_1$ gives rise to two disconnected wave packets are emerged into a single one at large time.
 Therefore, as in the Lipkin model, the ESQPT in the coupled top model can also be identified 
 through the particular dynamics of the Husimi function.
 
 More evident signatures of ESQPT are revealed in the features 
 of long-time averaged Husimi function Eq.~(\ref{AvHf}).
 For the coupled top model, it can be written as
 \be
   \ov{Q}_{ct}(p_1,q_1)=\la\zeta(p_1,q_1)|\ov{\rho}_1(\xi_1)|\zeta(p_1,q_1)\ra,
 \ee
 where $\ov{\rho}_1(\xi_1)=\mathrm{Tr}_2[\ov{\rho}(\xi_1)]$ with
 \be
   \ov{\rho}(\xi_1)=\lim_{T\to\infty}\frac{1}{T}\int_0^T\rho_t(\xi_1)dt. \notag
 \ee
 In the eigenstates of the post-quench Hamiltonian $H_{ct}(\xi_1)$, denoted by $\{|E_n\ra\}$, it
 is then straight to find that $\ov{Q}_{ct}(p_1,q_1)$ can be calculated as
 \begin{align}
   \ov{Q}_{ct}&(p_1,q_1) \notag \\
   &=\sum_n|\la\Psi_0|E_n\ra|^2\la\zeta(p_1,q_1)|\rho_1^{(n)}(\xi_1)|\zeta(p_1,q_1)\ra,
 \end{align}
 where $\rho_1^{(n)}(\xi_1)=\mathrm{Tr}_2(|E_n\ra\la E_n|)$.
 As we found in the Lipkin model, $\ov{Q}_{ct}(p_1,q_1)$ of the coupled top model also depends on the
 transition probabilitites between the initial state and the $n$th eigenstate of $H_{ct}(\xi_1)$. 
 
 In Figs.~\ref{AvgctHf}(a)-\ref{AvgctHf}(c), we plot $\ov{Q}_{ct}(p_1,q_1)$ for different 
 values of $\xi_1$ with $\xi_0=3$ and $j=30$.
 With decreasing $\xi_1$, the Husimi function exhibits a remarkable change as soon as $\xi_1\leq\xi_1^c=1.5$. 
 The ESQPT at $\xi_1^c=1.5$ is clearly associated with a significant extension of the Husimi function in phase space.
 The dramatical changes of $\ov{Q}_{ct}(p_1,q_1)$ in phase space with decreasing $\xi_1$ are more visible in its marginal
 distributions, as depicted in panels (d) and (e) of Fig.~\ref{AvgctHf}.  
 Consequently, the ESQPT in the coupled top model is signified as
 the dramatical extension of the Husimi function in phase space, as observed in the Lipkin model.   
 
 To provide further insights into the phase space signatures of ESQPT in the coupled top model, we consider 
 the second moment and Wehrl entropy of $\ov{Q}_{ct}(p_1,q_1)$.
 In Figs.~\ref{MWct}(a) and \ref{MWct}(b), we plot, respectively, $\ov{M}_{2,ct}$ and 
 $\ov{W}_{ct}$ as a function of $\xi_1$ with $\xi_0=3$.
 The critical value of $\xi_1$ for $\xi_0=3$ is $\xi_1^c=1.5$.
 The dramatic change in the behaviors of $\ov{M}_{2,ct}$ and $\ov{W}_{ct}$ as $\xi_1$ 
 passes through its critical value are clearly visible.
 For $\xi_1<\xi_1^c=1.5$, $\ov{M}_{2,ct}$ ($\ov{W}_{ct}$) is fixed at some smallest (largest) value, 
 which decreases (increases) with increasing $j$, 
 indicating that the Husimi function has maximum extension in this phase.
 Contrasting with $\xi_1<\xi_1^c$, we observe
 $\ov{M}_{2,ct}$ ($\ov{W}_{ct}$) increases (decreases) as $\xi_1$ increases when $\xi>\xi_1^c$. 
 These results suggest that the 
 largest extension of the Husimi function in phase space can be considered as one of signatures of ESQPT. 
 We further find that $\ov{M}_{2,ct}$ follows a power law scaling of the form $\ov{M}_{2,ct}\sim j^{\nu_M}$ 
 with scaling exponent $\nu_M$ depends on the value of $\xi_1$, as seen in the inset of Fig.~\ref{MWct}(a).
 On the other hand, the Wherl entropy exhibits a logarithmic scaling $\ov{W}_{ct}\sim\nu_W\ln(j)$ with
 $\nu_W$ varies with $\xi_1$ [inset in Fig.~\ref{MWct}(b)].
 The dependences of $\nu_M$ and $\nu_W$ on $\xi_1$ are shown in Fig.~\ref{MWct}(c). 
 We see that, as a function of $\xi_1$, both $\nu_M$ and $\nu_W$ behave differently in the two phase.  
 In particular, a rapid decrease in $\nu_M$ and $\nu_W$ is clearly visible around the critical point.
 This leads us to identify the critical point of ESQPT as the location of the minima points in the derivatives of 
 $\nu_M$ and $\nu_W$ with respect to $\xi_1$.
 Our numerically estimated critical points, together with the analytical ones obtained from Eq.~(\ref{Crxi}) are
 plotted in Fig.~\ref{MWct}(d).
 A good agreement between the numerical and analytical results can be clearly seen.
  
  \begin{figure}
  \includegraphics[width=\columnwidth]{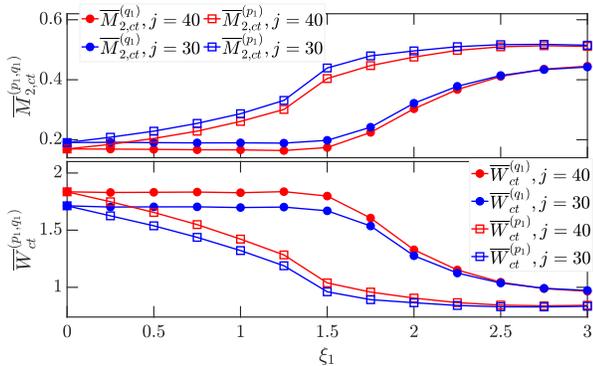}
  \caption{Second moment (upper panel) and Wehrl entropy (bottom panel) of 
  the marginal distributions of $\overline{Q}_{ct}(p_1,q_1)$ as a function of $\xi_1$ 
  for different $j$ with $\xi_0=3$.}
  \label{SMWct}
 \end{figure}
 
 In Fig.~\ref{SMWct}, we show the second moment and Wehrl entropy 
 of the marginal distributions of the Husimi function as a function of $\xi_1$ for different system size with $\xi_0=3$. 
 As expected, the behaviors of the marginal quantities change dramatically as the system 
 crossing of the critical point of ESQPT.
 Moreover, as observed in Lipkin model, the Husimi function of the coupled top model also exhibits 
 a larger degree of extension in the position direction.

 \section{Conclusions} \label{Fv}
 
 We have studied the phase space signatures of ESQPTs by means of Husimi function in two different models, namely
 Lipkin and coupled top model, both of them exhibit a second-order ESQPT 
 at certain critical energy. 
 We showed that the phase space signatures of ESQPT can be identified through 
 different properties of Husimi function and its marginal distributions. 
 We found that the different phases of ESQPT are revealed by distinct dynamical behaviors of the Husimi function
 and the particular dynamics of the Husimi function is able to detect the presence of ESQPT in both models.
 We also demonstrate that the long time average of the Husimi function 
 exhibits strikingly distinct features in different phases of ESQPT.
 The transition of the long time averaged Husimi function from two symmetrically localized wave packets to
 a single extended wave packet can be recognized as the main signature of ESQPTs in phase space. 
 To quantity the phase space spreading of the long time averaged Husimi function, 
 we further investigated the properties of the second moment 
 and Wherl entropy of the long time averaged Husimi function and its marginal distributions.  
 The singular features observed in their second moment and Wherl entropy 
 represent a visible manifestation of ESQPT.
 In turn, we employed these singular features to estimate the critical point of ESQPT and seen a 
 good agreement between the numerical estimations and the analytical results.
 
 Our findings confirm that phase space methods represents a powerful tool to understand ESQPTs 
 of many body systems, extending
 the previous works that focus on the phase space signatures of the ground state quantum phase transitions. 
 As ESQPTs studied in this work are quite general, we anticipate that the 
 ESQPTs in other systems, such as Rabi \cite{Puebla2016} and Dicke models \cite{Brandes2013,Magnani2014}, 
 will exhibit same signatures in phase space.
 It is an interesting future prospect to systematically explore the phase space signatures of ESQPTs 
 in various many body systems.  
 Another interesting extension of the present work would be to explore the phase space 
 signatures of the first-order ESQPTs, which
 characterized by the discontinuity of the density of states \cite{Stransky2016}. 
 Finally given the Husimi function has been measured in several experiments \cite{Eichler2011,Bohnet2016,Bouchard2017}, 
 and the realizations of the models studied in this work in quantum simulators 
 \cite{TianT2020,Zibold2010,Strobel2014,Hines2005}, 
 we expect that our results can be experimentally tested.

\acknowledgements
 
Q.~W. acknowledges support from the National Science Foundation of China under Grant No.~11805165, 
Zhejiang Provincial Nature Science Foundation under Grant No.~LY20A050001, and
Slovenian Research Agency (ARRS) under the Grant Nos.~J1-9112 and P1-0306.
This work has also been partially supported by the Consejer\'{i}a de Conocimiento, Investigaci\'{o}n y Universidad, Junta
de Andaluc\'{i}a and European Regional Development Fund (ERDF), ref.~SOMM17/6105/UGR and by the Ministerio de
Ciencia, Innovaci\'{o}n y Universidades (ref.~COOPB20364). FPB also thanks support from project UHU-1262561.
Computing resources supporting this work were partly provide by the CEAFMC and Universidad de Huelva 
High Performance Computer (HPC@UHU) located in the Campus Universitario el Carmen and funded by
FEDER/MINECO project UNHU-15CE-2848.

 \appendix
 
 \section{Critical point of ESQPT in the coupled top model} \label{AppA}
 
 In the semiclassical approach, the energy surface of the system 
 is the expectation value of the Hamiltonian in the coherent state.
 Therefore, the rescaled energy surface of the coupled top model is given by
 \be
    \mathcal{E}(p_1,q_1,p_2,q_2)=\la\Upsilon(\bp,\bq)|H_{ct}|\Upsilon(\bp,\bq)\ra/j,
 \ee
 where $|\Upsilon(\bp,\bq)\ra=|\zeta(p_1,q_1)\ra\otimes|\zeta(p_2,q_2)\ra$ with $\bp=(p_1,p_2)$ and $\bq=(q_1,q_2)$.
 By using the relations
 \begin{align}
    &\la\zeta|J_+|\zeta\ra=\frac{2j\zeta^\ast}{1+|\zeta|^2},\quad
    \la\zeta|J_-|\zeta\ra=\frac{2j\zeta}{1+|\zeta|^2}, \notag  \\
    &\la\zeta|J_z|\zeta\ra=j\left(\frac{|\zeta|^2-1}{|\zeta|^2+1}\right),
 \end{align}
 with $J_\pm=J_x\pm iJ_y$,
 it is straightforward to find that the rescaled energy surface of the coupled top model can be written as
 \begin{align}
    \mathcal{E}(\bp,&\bq)
       =\frac{1}{2}(p_1^2+q_1^2)+\frac{1}{2}(p_2^2+q_2^2)-2 \notag \\
       &+\frac{\xi}{4}q_1q_2\sqrt{4-(p_1^2+q_1^2)}\sqrt{4-(p_2^2+q_2^2)}.
 \end{align}
 The fixed points correspond to the values $(\bp_f,\bq_f)$ that produce the ground state energy of $H_{ct}$ are
 obtained by minimizing $\mathcal{E}(\bp,\bq)$ with respect to $\bp$ and $\bq$ 
 for a given value of $ \xi_0$. 
 The final results are given by
 \begin{align}
  (&\bp_f, \bq_f)=(p^f_1,p^f_2,q^f_1,q^f_2) \notag \\
  &=
  \begin{cases}
  (0,0,0,0)\quad &\text{for } \xi\leq1, \\
  (0,0,\pm\sqrt{\frac{2(\xi-1)}{\xi}},\mp\sqrt{\frac{2(\xi-1)}{\xi}})\quad &\text{for } \xi>1,
  \end{cases}
 \end{align}
 with energies
 \be
    \mathcal{E}_m=
    \begin{cases}
      -2 \quad &\text{for } \xi\leq1, \\
      -\left(\xi+\frac{1}{\xi}\right)\quad &\text{for } \xi>1.
    \end{cases}
 \ee
 
 For the ground state $|\Psi_0\ra$ of pre-quench $H_{ct}$ with $\xi=\xi_0>1$, 
 the energy of the post-quench Hamiltonian $H_{ct}(\xi_1)$ is given by
 \be
   \mathcal{E}(\xi_0,\xi_1)=\la\Psi_0|H_{ct}(\xi_1)|\Psi_0\ra.
 \ee
 As $|\Psi_0\ra=|\Upsilon(\bp_f,\bq_f)\ra$, the explicit expression of $\mathcal{E}(\xi_0, \xi_1)$ can be written as
 \be
   \mathcal{E}(\xi_0,\xi_1)=\frac{(\xi_0-1)[2\xi_0-\xi_1(\xi_0+1)]}{\xi_0^2}-2.
 \ee 
 For the critical quench $\xi_1^c$, we have $\mathcal{E}(\xi_0,\xi_1^c)=-2$.
 Hence, the critical quench $\xi_1^c$ is given by
 \be
    \xi_1^c=\frac{2\xi_0}{\xi_0+1},
 \ee
 with $\xi_0>1$.

%% Expanding the evolved state $|\psi_t\ra$ in the basis $\{|m_1,m_2\ra\}$ with $m_1,m_2\in\{-j,\ldots,j-1,j\}$, so that
%% $|\psi_t\ra=\sum_{m_1,m_2}C_{m_1,m_2}(t)|m_1,m_2\ra$ with $\sum_{m_1,m_2}|C_{m_1,m_2}(t)|^2=1$, 
%% the reduced density matrix of the first spin is
%% \be
%%    \rho_1(t)=\sum_{m_2}C_{m_1,m_2}(t)C^\ast_{m'_1,m_2}(t)|m_1\ra\la m'_1|.
%% \ee

 %%%%%%%%%%%%%%%%%%%%%%%%%%%%%%%%%%%%%%%%%%%%%%%
 %%%%%%%%%%%%%%%%%%%%%%%%%%%%%%%%%%%%%%%%%%%%%%%

%  \begin{figure}
%    \includegraphics[width=\columnwidth]{Figs/EsfCTP}
%  \caption{Contour plot of the classical energy surface for the coupled top model with $\lambda=0.4$ (a),
%  $\lambda=1$ (b), and $\lambda=1.4$ (c).}
%  \label{Cegcp}
% \end{figure}

\bibliographystyle{apsrev4-1}
\bibliography{PsESQPT}

\end{document}